# Diffusive and Ballistic Transport in Ultra-thin InSb Nanowire Devices Using a Few-layer-Graphene-AlO$_x$ Gate

Lior Shani[1*], Pim Lueb[2*], Gavin Menning[1], Mohit Gupta[1], Colin Riggert[1], Tyler Littman[1], Frey Hackbarth[1], Marco Rossi[2], Jason Jung[2], Ghada Badawy[2], Marcel A. Verheijen[2], Paul Crowell[1], Erik P. A. M. Bakkers[2], Vlad S. Pribiag[1&]

[1] *School of Physics and Astronomy, University of Minnesota, Minneapolis, Minnesota 55455, USA*

[2] *Department of Applied Physics, Eindhoven University of Technology, Eindhoven, The Netherlands.*

[*] Equal Contribution

[&] Corresponding Author: vpribiag@umn.edu

## Abstract

Quantum devices based on InSb nanowires (NWs) are a prime candidate system for realizing and exploring topologically-protected quantum states and for electrically-controlled spin-based qubits. The influence of disorder on achieving reliable topological regimes has been studied theoretically, highlighting the importance of optimizing both growth and nanofabrication. In this work we investigate both aspects. We developed InSb nanowires with ultra-thin diameters, as well as a new gating approach, involving few-layer graphene (FLG) and Atomic Layer Deposition (ALD)-grown AlO$_x$. Low-temperature electronic transport measurements of these devices reveal conductance plateaus and Fabry-Pérot interference, evidencing phase-coherent transport in the regime of few quantum modes. The approaches developed in this work could help mitigate the role of material and fabrication-induced disorder in semiconductor-based quantum devices.

## Introduction

Improving the electronic cleanliness of quantum devices based on low-dimensional semiconductors, such as nanowires or quantum wells, is of paramount importance for creating robust and tunable quantum states and for enabling quantum technologies. InSb and InAs NWs have a wide range of desirable properties, including strong spin-orbit[1] coupling (SOC) and large g-factors[2,3], that make their use appealing for realizing qubits based on Majorana Zero Modes (MZM)[4–7] and electrically-controlled spin-based qubits[8–10]. MZMs, which could be used to develop topologically-protected qubits, require proximitizing of nanowires with a conventional superconductor, such as Al. A key precursor of MZMs is the spin-helical state, in which momentum and spin become correlated. This arises in the normal state, without superconductors, and requires spin-orbit coupling and robust ballistic transport. However, establishing clean ballistic transport and the presence of the helical state have been sidestepped in most recent experiments on MZMs, which have prioritized more complex devices involving superconductors even when these underlying requirements were not reliably confirmed.

Recent theoretical analysis[11,12] suggests that disorder, due to inhomogeneous dielectric environments, surface charges on the nanowire and impurities in the crystal lattice that occur during the growth process, is a key factor in determining whether the topological regime of MZMs can be reliably achieved. The presence of quantized conductance is a clear indication of whether a nanowire device is sufficiently clean to host helical states and hence, potentially, MZMs, making investigations of quantized transport a critical first step towards ascertaining that MZMs can be realized using a given type of nanowire and nanofabrication process. Therefore, optimizing the materials and nanofabrication is an important task that is necessary in order to align the quality of quantum devices with the requirements of MZMs.

Past progress in InSb NW growth has produced high quality wires with a typical diameter of $90 - 120\ nm$[13]. However, the diameter of the wires was still 3-4 times larger than the Fermi wavelength[14] of InSb is ($\lambda_F \sim 30\ nm$). Therefore, reducing the diameter of the wire may unravel new phenomena by further enhancing quantum confinement, leading to larger inter-subband energy splitting and potentially reduced scattering in the relevant regime of few occupied transverse quantum modes.

Recent advances in the development of 2D materials offer new ways to electrostatically gate quantum devices by taking advantage of the low surface roughness and flexibility in stacking 2D materials. Owing to its high electrical conductivity, graphene has emerged as a widely studied material for the development of new, high-performance nanoelectronics devices such as sensors and transistors[15,16]. However, because monolayer graphene is a Dirac semimetal[17], external gating is required to achieve a 'metallic' electrical conductivity[18]. In addition, the surface of graphene is chemically inactive[19], which inhibits the growth of thin dielectrics using conventional techniques such as Physical Vapor Deposition (PVD) and Atomic Layer Deposition (ALD), thus restricting the usefulness of graphene as an electrostatic gate for tuning quantum devices. A possible solution for the issues above is using a graphene multilayer, since it is intrinsically metallic and has a more chemically active surface.

In this work we demonstrate the structural and electrical characterization of ultra-thin $\sim 10\ \mu m$ long InSb nanowires with a diameter of $\sim 50\ nm$. To optimize the dielectric environment surrounding the wire we developed a facile fabrication technique based on Few-Layer-Graphene (FLG) and ALD-grown AlOx to electrostatically gate the nanowires.

Experimental

The fabrication of stemless InSb NWs closely matches earlier techniques[20], where an InSb(111)B substrate with SiNx mask is used. However, differing from previously reported InSb NWs, the holes in the mask are created with reactive ion etching (RIE) (Oxford Plasmalab System 100) as opposed to wet chemical etching methods. The nanoholes in the mask are filled with gold droplets, allowing for growth using a metalorganic vapor-phase epitaxy (MOVPE) technique. Precursors of trimethylindium and trimethylantimony are used, where under high temperature the organo-indium and organo-antimony bonds are broken, allowing the indium and antimony to interact with the gold droplet. The gold acts as the catalyst particle for the vapor-liquid-solid (VLS) method[21].

The fabrication of the FLG-AlOx back gate begins with depositing FLG to a pre-patterned chip with gold rectangles. FLG was exfoliated from Highly Oriented Pyrolytic Graphite (HOPG, HQ Graphene) and positioned deterministically using a transfer station. The FLG flake is placed on SiO2 substrate and partially overlapping with a gold rectangle to connect the FLG to the voltage source. The typical areas of the FLG flakes were selected to be larger than $50\ \mu m^2$ to have a sufficient overlap with the gold rectangle and to contain the entire length of the wire. A $\sim 55\ nm$ of AlOx was deposited on the entire chip using Plasma Enhanced Atomic Layer Deposition (PE-ALD, Fiji 2, Ultratech) at $150\ C$. While the FLG surface is more reactive compared to monolayer graphene, nonetheless, it still lacks dangling bonds to react with the precursor using only thermal treatment, which causes the dielectric to grow with pinholes. Therefore, additional surface treatment is necessary to create nucleation sites for the precursor[22,23]. The PE-ALD allows deposition of dielectric layer on the FLG without pretreatment because it creates local defects using plasma that serves as nucleation site for the precursor.

To investigate the electrical properties of the NWs we used a mechanical transfer station with a micro manipulator and an optical microscope to position the NWs on the FLG-AlOx back gates, then patterned leads using conventional electron-beam lithography (see *SI*). The NW is coated with a layer of native oxide, therefore, to make contact to the wire it is necessary to remove the oxide before putting on the metallic leads. The native oxide was removed using Ar ion milling, followed by in-situ evaporation of $10\ nm$ of Ti and $140\ nm$ of Au. The devices were pumped overnight and measured in a dilution refrigerator at base indicated temperature of $\sim 10\ mK$ (unless indicated otherwise).

After the device has been electrically characterized, a transmission electron microscopy (TEM) lamella is prepared using a FEI Nova Nanolab 600i. Protective layers of carbon and platinum are deposited with an electron-beam in vacuum. A lamella is cut using a Gallium Focused Ion Beam (FIB) and transferred to a half-moon TEM grid. The lamella is thinned by FIB milling in steps at $30\ kV$, $16\ kV$ and finally $5\ kV$, which creates a window thinner than $100\ nm$. This window is subsequently studied with TEM, a probe-corrected JEOL ARM 200F.

**Results and discussion**

The first goal is to realize thin and long InSb wires and, as reported before[20], the size of the mask opening influences the resulting wire diameter. The challenge is to create small, but well-defined holes. Here, we found that RIE provides finer control of the dimensions of the opening in the mask than wet chemical etching (*see SI)*. By carefully tuning the growth temperature and incoming material flux (*SI)*, the dimensions of the wire are optimized while retaining a high yield. At sufficiently low overall flux the radial growth rate is 6x lower than the axial growth rate as a function of time (*SI*), resulting in high aspect ratio nanowires. This can be understood by the fact that radial growth is driven by a vapor-solid (VS) mechanism, and axial growth by the VLS mechanism in which growth is catalyzed by the Au particle. By optimizing growth parameters, a yield of $90 + \%$ has been obtained, widths in the range $50 - 60\ nm$ and length of $14\ \mu m +$ (Figure 1a and 1b).

Figure 1c shows a false color SEM image of a typical NW-on-FLG-AlOx-back-gate device. The blue area is the location of the deposited FLG. The entire chip is coated with conformal AlOx. We also fabricated NW devices, on a different chip, with a Ti/Au-AlOx back gate (see *SI*). The electrodes were placed on top of the ultra-thin InSb NW, shown in gold and green respectively. The surface underneath the NW, FLG coated with AlOx, appears to be much smoother than the Ti/Au leads. AFM measurements confirm that the surface of the FLG-AlOx is significantly smoother compared to the Ti/Au-AlOx with typical roughness of $\sim 170 \pm 70\ pm$ and $\sim 860 \pm 70\ pm$ respectively.

The cross-sectional TEM image (Figure 1d) shows a cut of the measured nanowire device, reported in all of the measured data. The wire has a hexagonal shape and width of $\sim 52\ nm$. The wire is capped with $\sim 3\ nm$ oxide, and no crystal defects have been observed in the wire. In addition, the TEM cross-section of the FLG-AlOx back gate revealed that the thickness of the FLG is $\sim 4.5\ nm$, see SI.

Because the NWs are very long, we are able to fabricate multiple devices on the same NW. In particular, we fabricated on the same NW a 'long' device with source-drain spacing $d_{sd} \approx 6\ \mu m$ much larger than the mean-free-path ($l_s \approx 300\ nm$) and a 'short' device with $d_{sd} \sim 200\ nm$, on the order of $l_s$. The long device provides insight into diffusive transport, while the short device gives insight into quasi-ballistic transport. The inset to Figure 1c shows a conductance plot as a function of gate voltage (*V<sub>g</sub>*) of the ultra-thin NW device with contact spacing of $\sim 6\mu m$. The conductance increases as a function of gate voltage up to the point of saturation at $\sim 0.3\ G$. Measuring the pinch-off and conductance of the ultra-thin NW with contact spacing of $\sim 6\mu m$ allows us to extract the field-effect mobility value, which is a property of the diffusive transport regime.

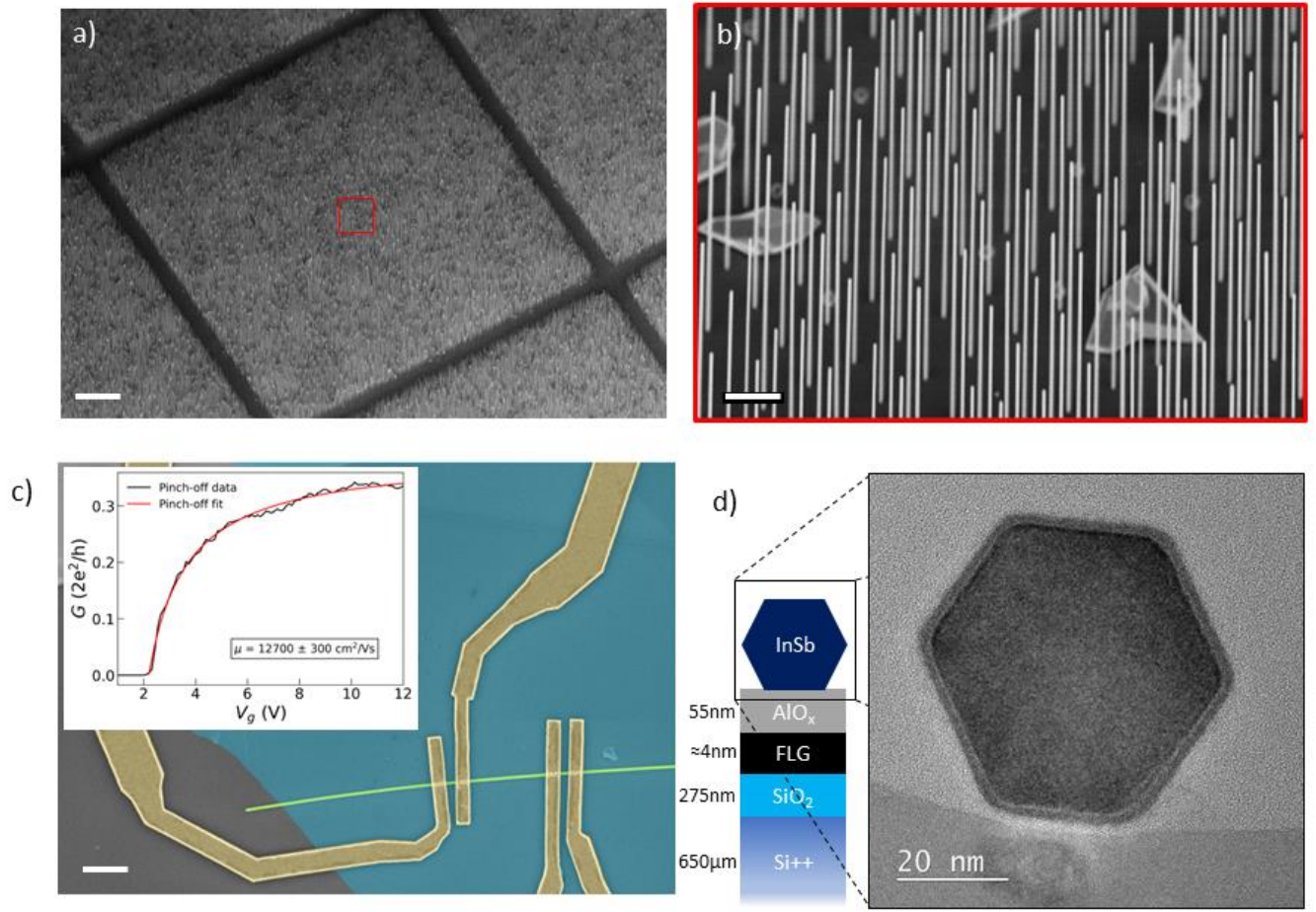

Figure 1. a) SEM image of a NW field taken under a 30°, where the scalebar is $30\ \mu m$. b) Close up of the NW field, showing high aspect ratio NWs. The larger objects appearing behind the NWs are parasitic growth on the mask. Scale bar is $1\ \mu m$. c) SEM image of a NW device placed on FLG-AlOx back gate, scale bar is $1\ \mu m$. The inset to the figure shows the transport data for the device with the $6\ \mu m$ spacing, the black line is the measured data, and the red line is the fitting to eq.1. d) Cross-sectional TEM overview of the measured NW device, showing defect free hexagonal crystal structuring having a width of $\sim 52\ nm$ and native oxide layer of $\sim 3\ nm$.

To extract the field-effect mobility, $\mu$, from this pinch-off data we treat the wire as a variable-conductance channel coupled to the gate electrode with a capacitance $C$. Additionally, we consider this channel in series with a fixed resistance $R_S$, which includes the contact resistance of the normal metal/wire interfaces, as well as the resistance of the fridge lines, filters, and measurement apparatuses. Together, this gives the gate-dependent conductance of the device as,

$$(1) \quad G(V_g) = \left(R_S + \frac{L^2}{\mu C(V_g - V_{th})}\right)^{-1}$$

where $L$ is the channel length and $V_{th}$ is the threshold gate voltage at which the Fermi level enters the conduction band and the device becomes conductive[13].

In order to fit this equation to our data, we calculate the capacitance between our gate and wire, using a self-consistent, finite-element Schrödinger-Poisson solver to simulate the exact geometry of our device, as described in Ref [24]. For the ultra-thin 6 $\mu m$ long channel, we calculate this capacitance to be 588 $aF$. Fitting the data in Fig 1c using this capacitance, we find our mobility to be $\mu = 12{,}700 \pm 300 \; cm^2/Vs$.

Figure 2a shows the differential conductance ($dI/dV$) map as a function of gate voltage ($V_g$) of the short device. The conductance map as function of $V_g$ shows quasi-periodic diamond shape patterns. A series of line cuts taken from the conductance map (Figure 2a) for $-1.95mV < V_{sd} < 1.95mV$ as a function of $V_g$ (shown in Figure 2b) demonstrate that the oscillations persist over a substantial range of gate voltages.

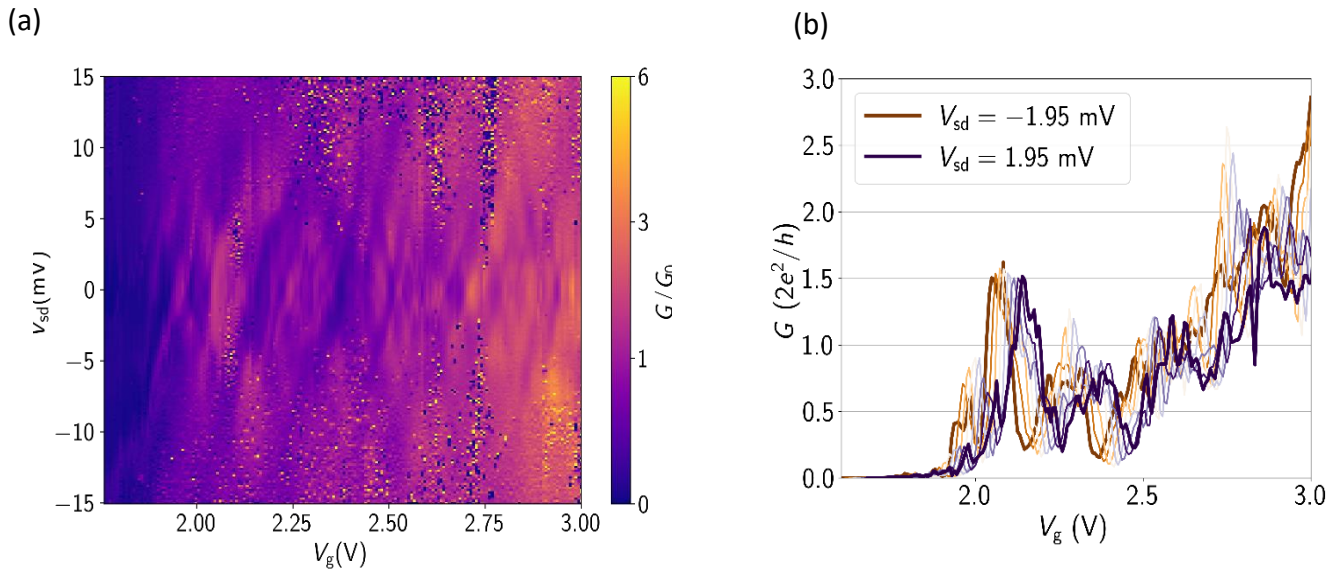

Figure 2. a) Differential conductance ($dI/dV$) vs $V_{sd}$ and $V_g$ of the device with contact spacing of 200 $nm$ at base temperature showing diamond shape pattern i.e., an indication of the FB interference. b) A series of line cuts from the 2D differential conductance map (Fig. 2a) between $V_{sd} = -1.95mV$ and $V_{sd} = 1.95mV$ as a function of $V_g$, purple and brown curve respectively.

The appearance of the diamond shape pattern indicates that the device acts as a Fabry−Pérot (FP) interferometer, a partially transmitting cavity for electron waves. The reflection may occur in part as a result of band bending near the contacts, as expected for a metal-semiconductor interface, and was observed before in InSb and InAs nanowires[25–27]. In addition, metallic contacts screen the electric field, such that the back-gate tunability on the channel close to the contacts is reduced, resulting in different electron densities there vs. in-between the contacts. As a result of these barriers, electrons experience multiple partial reflections near the contacts while propagating phase-coherently, which gives rise to FP interference, setting a lower bound of $200\ nm$ for the length over which phase-coherent and quasi-ballistic transport can be achieved in these devices. The FP oscillations are very well defined, with a large relative conductance modulation suggestive of transport that is close to the 1D limit. Though revealing phase-coherent transport, FP oscillations may obscure other quantum phenomena, in particular quantized conductance plateaus associated with individual transverse quantum modes.

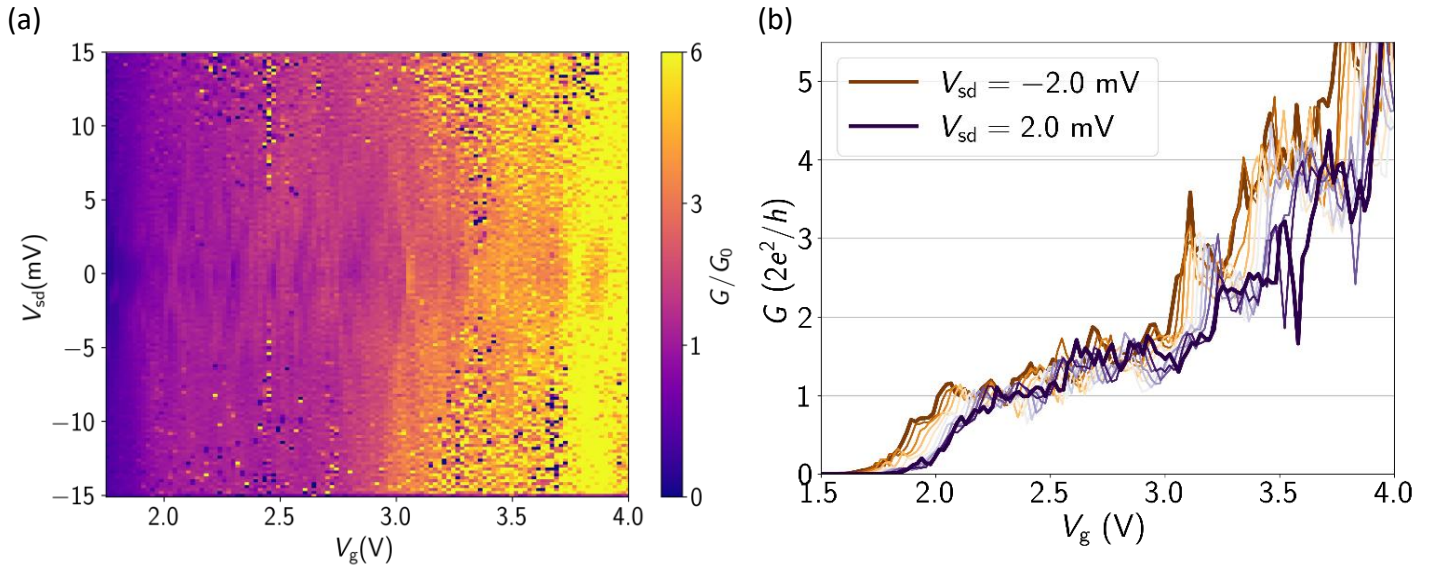

Figure 3. a) Differential conductance $(dI/dV)$ vs $V_{sd}$ and $V_g$ of the device with contact spacing of $200\ nm$ under $2.8\ T$ out-of-plane magnetic field showing a suppression of FB interference. b) A series of line cuts from the 2D differential conductance map at $B = 2.8T$ (Fig 3a) between $V_{sd} = -2\ mV$ and $V_{sd} = 2\ mV$ as a function of $V_g$, purple and brown curve respectively. The curves show the emergence of a plateau at $1\ G_0$, and a second plateau near $1.5\ G_0$ compatible with Zeeman splitting due to the applied B-field.

Applying an external magnetic field to the ultra-thin NW lifts the spin degeneracy and may cause suppression of FB oscillation by inducing dephasing between multiple trajectories. To investigate the presence of conductance plateaus, we apply an out-of-plane magnetic field ($B = 2.8T$) to suppress the FP oscillations (Figure 3). Figure 3b shows a series of line cuts taken at $-2mV < V_{sd} < 2mV$ from the differential conductance map, confirming that the FP oscillations are substantially reduced with respect to the $B = 0$ data. The line cuts reveal a conductance plateau that remains close to $G_0$ over a wide range of source-drain bias values, compatible with quantized transport superimposed with some residual oscillations due to FP interference.

In addition, a further plateau-like feature appears near 1.5 $G_0$, consistent with Zeeman-split subbands. At more positive gate voltages conductance oscillations, possibly due to FP interference, prevent the observation of any other clear plateaus. However, for MZM experiments, identification of the lowest one or two subbands is generally sufficient.

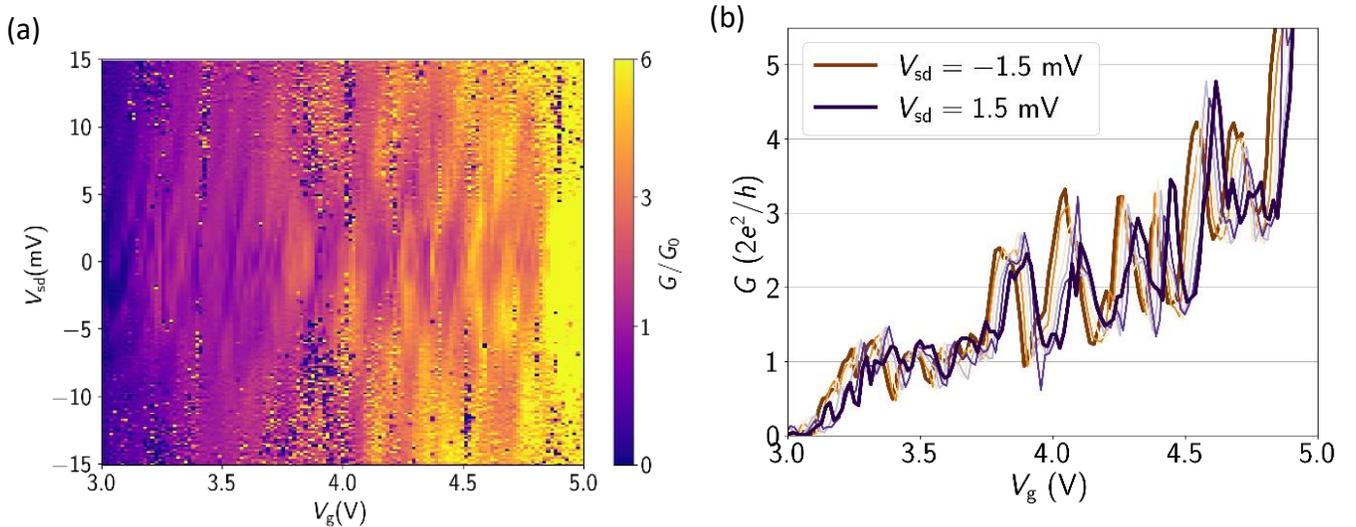

Figure 4. a) Differential conductance ($dI/dV$) vs $V_{sd}$ and $V_g$ of the device with contact spacing of 200 $nm$ at 400 $mK$ showing a suppression of FB interference. b) A series of line cuts from the 2D differential conductance map (Fig 4a) between $V_{sd} = -1.5\ mV$ and $V_{sd} = 1.5\ mV$ as a function of $V_g$, purple and brown curve respectively. The curves show the emergence of a plateau at 1 $G_0$ thus highlight the fact the FB interference can suppress potential conductance plateaus.

To further investigate the robustness of the observed 1 $G_0$ plateau, we measure conductance at a higher temperature, which is expected to reduce the phase coherence and thus further suppress resonances such as those due to FP interference. Figure 4a shows a differential conductance map as function of $V_g$ and $V_{sd}$ taken at 400 $mK$ and $B = 0$. Note that the pinch-off voltage increased to $V_g \sim 3V$ after the temperature change, possibility due to a spontaneous charge switch in the dielectric environment, either in the native oxide shell surrounding the wire or the ALD deposited gate dielectric. The finer oscillatory features seen at $10\ mK$ in Figure 2a are suppressed at $400\ mK$, however clear FP interference is still present, as evidenced by the diamonds in the differential conductance map. Figure 4b shows a series of line cuts for $-1.5 < V_{sd} < 1.5\ mV$. The oscillations appear to contain fewer harmonic components than in the $10\ mK$ data, consistent with suppression of phase coherence and hence interference between fewer coherent trajectories. For $3.2 < V_g < 3.8\ V$ the conductance values again coalesce near the value of $1\ G_0$, providing further evidence that this plateau feature is likely due to the lowest transverse subband in the nanowire.

In order to gain better understanding of the origin of the FB oscillations and their interplay with subband quantization, we performed tight-binding numerical simulations using the KWANT[28] Python software package. We show that one possible origin of FP oscillations in our system is the difference in the chemical potential between the metallic leads and the semi-conducting nanowire. The Hamiltonian of our nanowire is defined as:

$$(2) \quad H = (\frac{p^2}{2m^*} - \mu) - i\alpha_R \sigma_z \frac{\partial}{\partial x} - \frac{g\mu_B}{2}\vec{\sigma}\cdot\vec{B}$$

The first term of the Hamiltonian is the kinetic energy, the second term of the Hamiltonian describes the Rashba SOC, where $\alpha_R$ is the Rashba spin-orbit coefficient and $\sigma_z$ is the z-component of the Pauli matrix, and the third term is the Zeeman coupling due to an applied magnetic field $\vec{B}$.

We simulate this by discretizing $H$ over a square mesh with a lattice constant $a = 5\ nm$. The width of nanowire is set to $60\ nm$ with a contact spacing of $200\ nm$[29]. The Hamiltonian of the leads includes only the kinetic energy term from the above defined $H$. Additional details regarding the simulation parameters are provided in the *SI*.

Figure 5a shows the subbands for a one-dimensional semiconductor with spin-orbit coupling and Zeeman coupling. The subbands are shifted relative to each-other along the k-axis due to the SOC. The red curve in Figure 5b shows conductance as a function of energy through this system. This is valid under the assumption that there is no chemical potential difference between leads and the wire and we obtain clean quantized conductance plateaus. However, this does not represent a typical NW experiment because the chemical potential of the NW is tuned by applying electric field using a gate, and the field in the leads area is screened by the metallic contacts. This gives rise to a non-uniform potential profile along the length of the wire. To account for this we implement the gate potential as a hyperbolic tangent function as described in Ref.[30]. This can result in FP oscillations in pristine wires without any defects. The blue curve in Figure 5b shows the conductance, $G$, in units of $G_0$ as a function of gate voltage. The curve shows conductance oscillations, making the previously visible plateaus obscure. The results obtained from our simulations agree qualitatively with the data in Figure 2b.

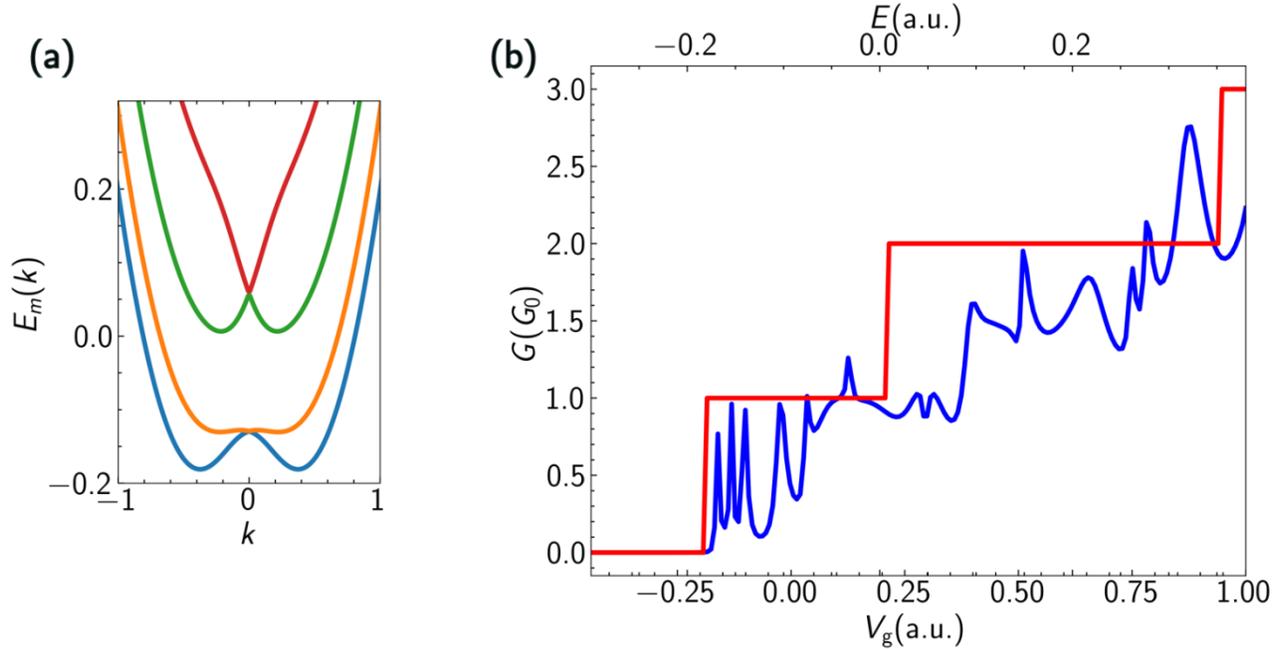

Figure 5. a) subband in a semi-conducting nanowire with spin-orbit interaction and Zeeman coupling. b) The red curve shows an energy sweep along the subbands of the wire. The y-axis is $G$ in the units of $G_0$ and the x-axis (upper) is a normalized energy scale in arbitrary units. The blue curve shows the conductance as a function of gate voltage under non-uniform gate potential. The Y-axis is $G$ in the units of $G_0$ as a function and the X-axis (lower) is the gate voltage in arbitrary units.

In conclusion, we fabricated $\sim 10\,\mu m$ long and $\sim 55\,nm$ wide InSb nanowires. We used Few-Layer-Graphene as the conducting layer for local back gate, yielding reduced surface roughness relative to Ti/Au backgates. The electrical properties of the ultra-thin InSb nanowires with FLG-AlO$_x$ local back gate were characterized at $mK$ temperatures. By leveraging the large length of the nanowires ($\sim 10\,\mu m$) we explored electronic transport in the diffusive transport regime on the same wire on which the ballistic regime was also accessed, the first with contact spacing of $\sim 6\,\mu m$ and the second with contact spacing of $\sim 200\,nm$. In the ballistic regime, the devices reveal Fabry-Pérot interference due to multiple partial reflections of the phase-coherently propagating electron waves. The visibility of the Fabry-Pérot oscillations competes with that of conductance plateaus, yet the plateaus remain detectable, suggesting that for the channel length of $\sim 200\,nm$ transport occurs phase-coherently and ballistically through a few discrete quantum modes. Quantized conductance at zero magnetic field has been very difficult to observe in an InSb nanowire devices[31] and remains a key test for a device's suitability for MZM experiments. Our results suggest that reducing the nanowire diameter, along with optimizing gate materials could be a viable path towards realizing MZMs.

**Acknowledgements**

This work was supported by the Department of Energy under Award No. DE-SC0019274. Portions of this work were conducted in the Minnesota Nano Center, which is supported by the National Science Foundation through the National Nanotechnology Coordinated Infrastructure (NNCI) under Award Number ECCS-2025124. Parts of this work were carried out in the Characterization Facility, University of Minnesota, which receives partial support from the NSF through the MRSEC (Award Number DMR-2011401) and the NNCI (Award Number ECCS-2025124) programs. This work was supported by the European Research Council (ERC TOCINA 834290). The authors recognize Solliance, a solar energy R&D initiative of ECN, TNO, Holst, TU/e, IMEC and Forschungszentrum Jülich, and the Dutch province of Noord-Brabant for funding the TEM facility. TU/e acknowledges the research program "Materials for the Quantum Age" (QuMat) for financial support. This program (registration number 024.005.006) is part of the Gravitation program financed by the Dutch Ministry of Education, Culture and Science (OCW).

**References**

[1] I. Van Weperen, B. Tarasinski, D. Eeltink, V.S. Pribiag, S.R. Plissard, E.P.A.M. Bakkers, L.P. Kouwenhoven, and M. Wimmer, Phys. Rev. B - Condens. Matter Mater. Phys. **91**, 1 (2015).

[2] I. Van Weperen, S.R. Plissard, E.P.A.M. Bakkers, S.M. Frolov, and L.P. Kouwenhoven,

Nano Lett. **13**, 387 (2013).

[3] S. Nadj-Perge, S.M. Frolov, J.W.W. Van Tilburg, J. Danon, Y. V. Nazarov, R. Algra, E.P.A.M. Bakkers, and L.P. Kouwenhoven, Phys. Rev. B - Condens. Matter Mater. Phys. **81**, 2 (2010).

[4] Y. Oreg, G. Refael, and F. Von Oppen, **177002**, 1 (2010).

[5] R.M. Lutchyn, J.D. Sau, and S. Das Sarma, Phys. Rev. Lett. **105**, 1 (2010).

[6] V. Mourik, K. Zuo, S.M. Frolov, S.R. Plissard, E.P.A.M. Bakkers, and L.P. Kouwenhoven, Science . **336**, 1003 (2012).

[7] S.M. Albrecht, A.P. Higginbotham, M. Madsen, F. Kuemmeth, T.S. Jespersen, J. Nygård, P. Krogstrup, and C.M. Marcus, Nature **531**, 206 (2016).

[8] V.S. Pribiag, S. Nadj-Perge, S.M. Frolov, J.W.G. Van Den Berg, I. Van Weperen, S.R. Plissard, E.P.A.M. Bakkers, and L.P. Kouwenhoven, Nat. Nanotechnol. **8**, 170 (2013).

[9] J.W.G. Van Den Berg, S. Nadj-Perge, V.S. Pribiag, S.R. Plissard, E.P.A.M. Bakkers, S.M. Frolov, and L.P. Kouwenhoven, Phys. Rev. Lett. **110**, 1 (2013).

[10] S. Nadj-Perge, S.M. Frolov, E.P.A.M. Bakkers, and L.P. Kouwenhoven, Nature **468**, 1084 (2010).

[11] S. Ahn, H. Pan, B. Woods, T.D. Stanescu, and S. Das Sarma, Phys. Rev. Mater. **5**, 1 (2021).

[12] B.D. Woods, S. Das Sarma, and T.D. Stanescu, Phys. Rev. Appl. **16**, 1 (2021).

[13] Ö. Gül, D.J. Van Woerkom, I. Van Weperen, D. Car, S.R. Plissard, E.P.A.M. Bakkers, and L.P. Kouwenhoven, Nanotechnology **26**, (2015).

[14] Y. Chen, S. Huang, D. Pan, J. Xue, L. Zhang, J. Zhao, and H.Q. Xu, Npj 2D Mater. Appl. **5**, 1 (2021).

[15] B. Braunecker and P. Simon, Phys. Rev. Lett. **111**, 1 (2013).

[16] J.A. Alexander-Webber, A.A. Sagade, A.I. Aria, Z.A. Van Veldhoven, P. Braeuninger-Weimer, R. Wang, A. Cabrero-Vilatela, M.B. Martin, J. Sui, M.R. Connolly, and S. Hofmann, 2D Mater. **4**, (2017).

[17] K.S. Novoselov, A.K. Geim, S. V. Morozov, D. Jiang, M.I. Katsnelson, I. V. Grigorieva, S. V. Dubonos, and A.A. Firsov, Nature **438**, 197 (2005).

[18] L. Liao and X. Duan, Mater. Sci. Eng. R Reports **70**, 354 (2010).

[19] A. Chouhan, H.P. Mungse, and O.P. Khatri, Adv. Colloid Interface Sci. **283**, 102215
13

(2020).

[20] G. Badawy, S. Gazibegovic, F. Borsoi, S. Heedt, C.A. Wang, S. Koelling, M.A. Verheijen, L.P. Kouwenhoven, and E.P.A.M. Bakkers, Nano Lett. **19**, 3575 (2019).

[21] A. You, M. Be, and I. In, (2004).

[22] J.H. Jeon, S.K. Jerng, K. Akbar, and S.H. Chun, ACS Appl. Mater. Interfaces **8**, 29637 (2016).

[23] R.H.J. Vervuurt, W.M.M.E. Kessels, and A.A. Bol, Adv. Mater. Interfaces **4**, 1 (2017).

[24] D. Eeltink, MSc Thesis Delft University of Technology (2013).

[25] J. Kammhuber, M.C. Cassidy, F. Pei, M.P. Nowak, A. Vuik, O. Gül, D. Car, S.R. Plissard, E.P.A.M. Bakkers, M. Wimmer, and L.P. Kouwenhoven, Nat. Commun. **8**, 1 (2017).

[26] A. V. Kretinin, R. Popovitz-Biro, D. Mahalu, and H. Shtrikman, Nano Lett. **10**, 3439 (2010).

[27] Z. Yang, B. Heischmidt, S. Gazibegovic, G. Badawy, D. Car, P.A. Crowell, E.P.A.M. Bakkers, and V.S. Pribiag, Nano Lett. **20**, 3232 (2020).

[28] C.W. Groth, M. Wimmer, A.R. Akhmerov, and X. Waintal, New J. Phys. **16**, (2014).

[29] Z. Yang, P.A. Crowell, and V.S. Pribiag, **1**, 1 (2021).

[30] D. Rainis and D. Loss, Phys. Rev. B - Condens. Matter Mater. Phys. **90**, 1 (2014).

[31] J. Kammhuber, M.C. Cassidy, H. Zhang, Ö. Gül, F. Pei, M.W.A. De Moor, B. Nijholt, K. Watanabe, T. Taniguchi, D. Car, S.R. Plissard, E.P.A.M. Bakkers, and L.P. Kouwenhoven, Nano Lett. **16**, 3482 (2016).